\documentclass[runningheads]{llncs}
\usepackage{graphicx}
\usepackage{listings}
\usepackage{multirow}
\usepackage{makecell}
\usepackage{multirow}
\usepackage{xcolor}
\usepackage[hidelinks]{hyperref}
\usepackage[utf8]{inputenc}
\usepackage[small]{caption}
\usepackage[hyphenbreaks]{breakurl}

\begin{document}
\title{ID-based self-encryption via Hyperledger Fabric based smart contract}
%
%
\author{Ilya Grishkov\inst{1} \and
Roland Kromes\inst{1} \and
Thanassis Giannetsos\inst{2} \and
Kaitai Liang\inst{1}}
\authorrunning{I. Grishkov et al.}
%
\institute{Cyber Security Group, Delft University of Technology, The Netherlands 
\email{I.Grishkov-1@student.tudelft.nl, R.G.Kromes@tudelft.nl,
Kaitai.Liang@tudelft.nl}
\and Ubitech Ltd, Digital Security \& Trusted Computing Group, Athens, Greece
\email{agiannetsos@ubitech.eu}
}
\maketitle              
\begin{abstract}
This paper offers a prototype of a Hyperledger Fabric-IPFS based network architecture including a smart contract based encryption scheme that meant to improve the security of user's data that is being uploaded to the distributed ledger. A new extension to the self-encryption scheme was deployed by integrating data owner's identity into the encryption process. Such integration allows to permanently preserve ownership of the original file and link it to the person/entity who originally uploaded it. Moreover, self-encryption provides strong security guarantees that decryption of a file is computationally not feasible under the condition that the encrypted file and the key are safely stored.


\keywords{Blockchain  \and IPFS \and Self-Encryption \and Security \and Hyperledger Fabric}
\end{abstract}
\section{Introduction}
\label{sec:Introduction}

The modern world is increasingly adopting blockchain technology. The first major market adoption of blockchain happened in 2009 when Bitcoin was introduced \cite{nakamoto2008bitcoin}. Interest in blockchain solutions grew over the years and lead to the invention of Ethereum - Bitcoin peer but with support for smart contracts which are digital codes enabling the description of complete business logic. \cite{buterin2014ethereum}. The introduction of smart contracts leads to further development in the field of blockchain and created demand for more industry-friendly solutions that allow to identify users of the system (Know-Your-Customer, Anti-Money-Laundering). Hyperledger Fabric was then introduced as a highly modular permissioned blockchain that allows great customization to suit particular industrial needs \cite{cachin2016architecture}. Given its customizability and modularity, Hyperledger Fabric (HLF) is a perfect platform for extending it with various trust and privacy preservation solutions. 

According to Huang et al. \cite{8864988} the main component of a blockchain that is being attack the most is a smart contract. High frequency of attack on a component designed to handle user private data suggests a need for an alternative approach to handling sensitive information other than just sending it raw to the ledger. An local data encryption prior to sending data to the smart contract could be a solution.

Self-encryption was introduced as a mean of encrypting files that "requires no user intervention or passwords" \cite{irvine2010self}. This algorithm can be used for local encryption of files, encrypted chunks of which will be later uploaded to a cloud-based storage or to a distributed file system (e.g., IPFS \footnote{https://ipfs.io/}) Pointers to the encrypted chunks are then sent to the ledger. It can be noted that storing only the hash values of the encrypted data chunks on blockchain ledger is vital when the data size is significant.  The authors in \cite{ATrueDecentralizedImplementationBasedOnIoTandBlockchainaVehicleAccidentUseCase} point out that sending data to a blockchain frequently, when the data size is large, can cause the entire blockchain network to crash. Sending only the hash values of the given data is more optimal as a hash value is usually 32 bytes long. While this solution allows to keep file content private, the file itself is not linked in any way to its owner. A variant of identity based encryption can tackle this problem. If a file is self-encrypted with owners identity used during the encryption process, this file remains linked to the person who initially uploaded it to the blockchain. This way original ownership can be preserved. 

This paper aims at exploring trust and privacy preserving solutions in Hyperledger Fabric blockchain. More specifically the goal is to further investigate the utility of a combination of identity based encryption and self-encryption as means of improving security of the data in the HLF; extend the previously done research by \cite{park2022using} and implement ID-based self-encryption via Hyperledger Fabric smart contract. Hence the main research question is: "How can security of Hyperledger Fabric smart contracts be improved using ID-based self-encryption?"

Within this paper an approach of integrating ID-based self-encryption is presented. Moreover a detailed description of prototype implementation is given. In addition to this implementation of ID-based self-encryption, a practical fully decentralized network architecture for storing encrypted data has also been deployed. In this proposed network, the data owner can use ID-based self-encryption to store encrypted data in a decentralized and secure manner. The encrypted data chunks are stored in an InterPlanetary File System (IPFS) which is a decentralized systems for file storage. To store the references (hash values)  of the encrypted data chunks, the Hyperledger Fabric blockchain was used.\\ 

This work is structured as follows. Section \ref{sec:StateOfTheArt} describes related works used to achieve the goal. Section \ref{sect:methodology} gives a background about the the implementation of ID-based self-encryption. Section \ref{sect:ID-based self-encryption} discusses the inner workings and the proposed implantation of ID-based self-encryption. The performance and security analysis of the proposed implementation is presented in section \ref{sect:Results}.  Finally, the work is discussed in section \ref{sect:Discussion}, and concluded in section \ref{sect:Conclusion}.

\section{State of the Art}
\label{sec:StateOfTheArt}

Blockchain is a distributed ledger technology that provides immutability and transparency of data to members of the blockchain network \cite{BCchallengesAndOpportunitiesSurvey}. The blockchain is also a peer-to-peer network in which participants are known to each other. Authentication of participants is ensured using elliptic curve cryptography. Today's blockchains enable the deployment of complete business logic in a seamless manner. These digital business logic are also known as smart contracts.

Blockchain technology is used in several use cases such as smart city, smart agriculture, vehicle networks \cite{AdaptationOfAnEmbeddedArchitectureToRunHyperledgerSawtoothApplication} and also healthcare \cite{BCHealthcare}. In the latter two cases, the privacy and ownership of data transmitted from a driver and medical patient is particularly important, as this data may contain privacy sensitive information. It can be noted that in these latter use case using a private blockchain such as Hyperledger Fabric can be a more optimal choice as they can provide higher security and privacy level.

The topic of security and privacy of the Hyperledger Fabric has been thoroughly studied \cite{yamashita2019potential}, \cite{dabholkar2019ripping}, \cite{stamatellis2020privacy}. Moreover a research has been conducted this year by a student of Delft University of Technology \cite{park2022using} addressing similar issue of improving HLF security using self-encryption.

The concept of self-encryption was introduced by Yu Chen \cite{chen2009self}. The approach of the original paper involves converting a file into a bit stream, extract the key by randomly selecting bits from the stream and then doing the encryption using that key. After the encryption the key and the encrypted file should be stored separetely, e.g. the key can be stored locally, while the encrypted file can be sent to a server. 

The original encryption scheme was also extended by Moch Rezky Debby Rahardjo \cite{rahardjo2017design}. According to the paper, "The modification is located in dividing the plaintext and ciphertext into 1024-bit chunks at XOR process and using the date when encryption process starts as a seed. The modification also adds the database for the key management function". Storing the key and the encrypted chunks in separate places makes in computationally not feasible to get the original data.

The later industrial adaptation of the self-encryption scheme happened when a team lead by David Irvine made self-encryption the core of his company's (MaidSafe) product - SAFE Network \cite{irvine2010self}. Irvine's implementation of the self-encryption scheme will be the basis of this work, hence a more detailed explanation of the implementation of the algorithm will be given. 

Figure 1 shows the encryption process. First, the original file is getting split into minimum of 3 file chunks. After the file is split into chunks the algorithm creates a data map, where the key needed for decryption will be stored. Each chunk is then hashed and those hashes are written to the data map. Parts of those hashes are used as a key and initialization vector for AES 128 algorithm that encrypts each file chunk. When encryption is done, each encrypted chunk is obfuscated with the previously computed hash values by applying a XoR function. At the end of the process the encryption scheme returns a data map that is going to be later used for decryption, and the encrypted file chunks.

\begin{figure}[h]
\centering
  \includegraphics[width=11cm, height=8cm]{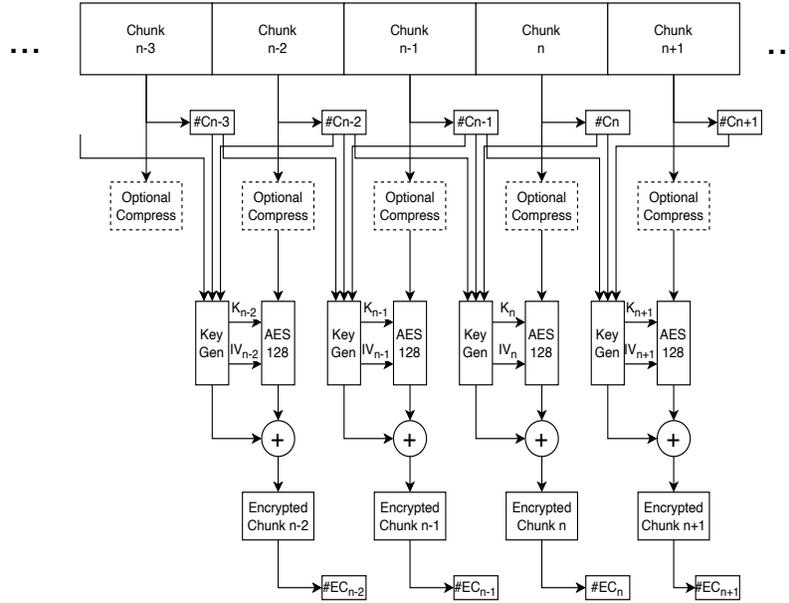}
  \caption{Self-encryption process, adopted from \cite{maidsafeselfencryption}}
  \label{fig:self_encryption}
\end{figure}

\section{Background}
\label{sect:methodology}
Original implementation of the self-encryption schema by David Irvine \cite{maidsafeselfencryption} was modified and used for this research. The use of rayon library (which adds parallelization to the code) was removed from the algorithm, due to the fact that  the compilation target (WebAssembly) only supports single-threaded code.  Additionally the code base was modified to include an interface for communication with the external code. Changes were also done to the Cargo.toml to make the code compatible with the target. Modified self-encryption algorithm was compiled to WebAssembly and run in a virtual machine (VM) and invoked from the code of the developed local application (which allows the interaction with the Hyperledger Fabric Smart Contract). A more detailed description of the process will be given in the section \ref{sect:ID-based self-encryption}. The benchmarks of this implementation will be provided in section \ref{sect:Results}.

Hyperledger Fabric test network v2.4.3 was used. Test network was deployed to Docker based on the tutorials provided by Hyperledger Fabric \footnote{Usage of the command requires navigating to the root directory of the test-network, provided by the Hyperledger Fabric \cite{tutorial}} .

Smart contract was then deployed (detailed in section \ref{sect:ID-based self-encryption}).

Encrypted file storage is handled by the IPFS, which is a distributed Torrent database, which uses hashes of files to address its content. IPFS node was also deployed to Docker. For IPFS deployment two directories (staging and data) were mounted on the host file system to persist the stored data, when the container is stopped. 
Hyperledger Fabric provides official software development kit (SDK) for 3 languages: Go, Java, Javascript. Go was chosen for implementation of the project, due to ease of integration with both Hyperledger Fabric and the IPFS. Encryption library is written in Rust and is compiled to WebAssembly, hence a way to call WebAssembly was needed. Go also provides support for Wasmer library that allows to call WebAssembly function directly from Go code. 

\section{ID-based self-encryption}
\label{sect:ID-based self-encryption}
\subsection{Integrating identity into the encryption}
This paper offers an extension to the algorithm proposed by Irvine \cite{irvine2010self}. Encryption step in the original algorithm is modified to include identity of a person who is running the algorithm into the encryption process. Instead of using part of the chunk hash as a key for AES 128, the result of XoR of the hashed identity and the chunk hash is used as a key. The identity can be any string of any length. If the length of this string is shorter than the length of the key, then the cycle function is applied to the string, which repeats the iterator of a string. The hashing function SipHash 1-3 is used to hash the identity of a user, before passing it to the XoR function. Figure 2 demonstrates the process of encrypting a file using the modified version of self-encryption with identity integrated into the encryption process.

\begin{figure}[h]
    \centering
  \includegraphics[width=11cm, height=8cm]{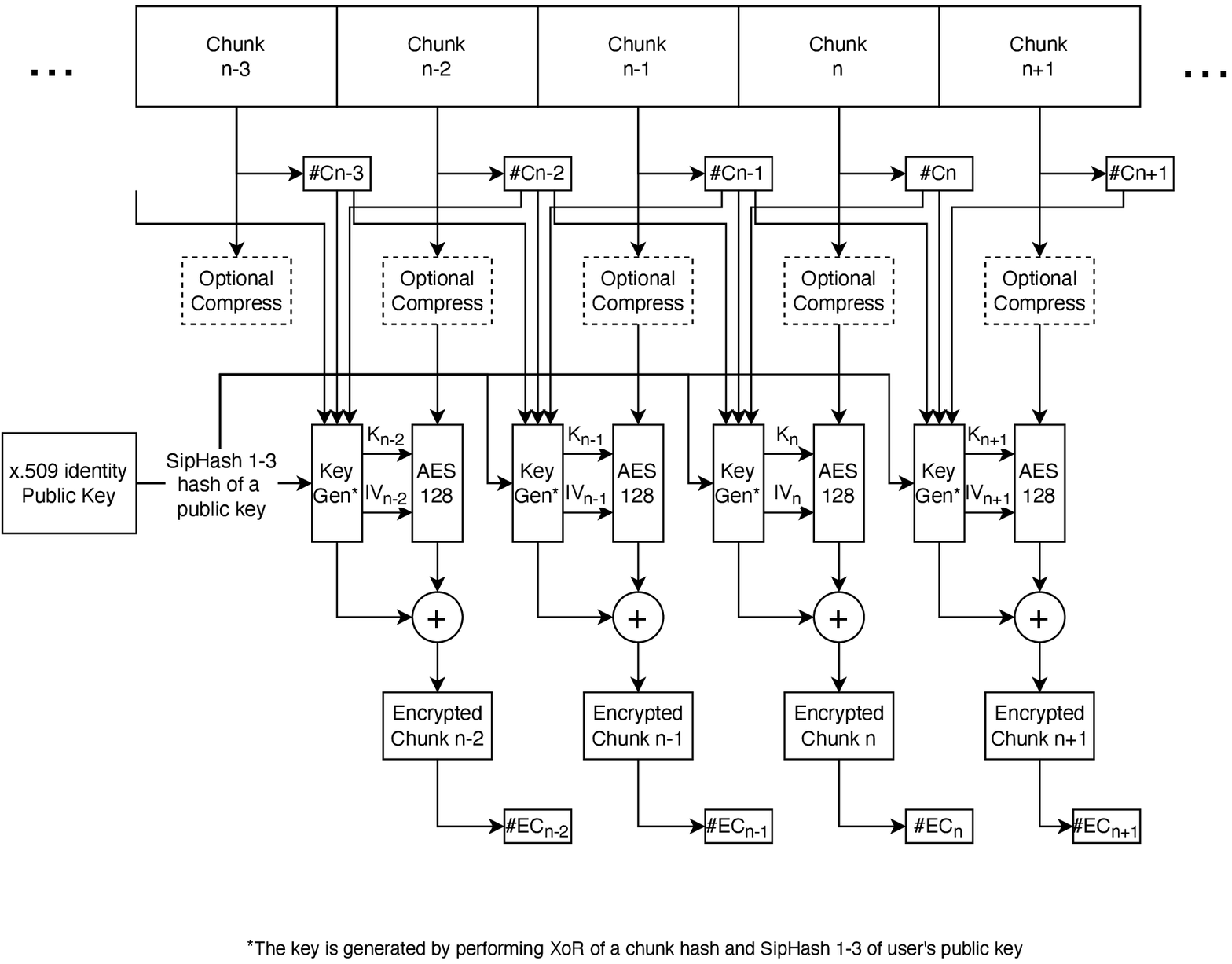}
  \caption{ID-based self-encryption process}
  \label{fig:id-based-self-encryption}
\end{figure}

Decryption of the file, that was encrypted using ID-based self-encryption, is similar to that of a regular self-encryption, with the key for AES 128 being the only different part. The decryption calculates the key the same way the encryption does it by applying XoR function to the hash of identity and the chunk hash from the data map.

The implementation of the encryption scheme can be found on GitHub \footnote{https://github.com/ilyagrishkov/ib-self-encryption-rust}

\subsection{Connecting the encryption algorithm and the local application}
The implementation of the identity-based self-encryption is written purely in Rust, while the client application is written in Go. This creates a demand for a way to integrate Rust library into Go code. Among the solutions to tackle the problem are:
\begin{enumerate}
    \item Use Go tools to assemble the Go code and compile Rust code into a static library. Then link compiled code using additional assembly "glue-code" \cite{filippo_valsorda_2019}.
    \item Compile Rust to a static library and call it from the Go code using Go build-in pseudo-library C for interacting with native interfaces.
    \item Compile Rust to WebAssembly (WASM) code and call it from Go using Wasmer library \footnote{https://wasmer.io/}. 
\end{enumerate}

All of the methods have been successfully tried. The first two methods do not allow cross compilation, because both of them require compiling Rust to a static library, which is platform-specific. Additionally, the first methods requires the use of assembly language, which is different on different processor architectures and operating systems. The second method also uses C pseudo-library, which does not allow cross-compilation of the Go code. Overall, both methods are very \emph{platform-specific}, which makes them less preferable choice.

The third method was chosen for connecting Rust library to Go code. Compiling Rust to WASM to use as a standalone application or a library, can be done using the following command:
\begin{lstlisting}[language=bash]
  $ cargo build --target=[chosen_target]
\end{lstlisting}

where \textit{chosen\_target} is a WebAssembly target that can be either wasm32-unknown-unknown or wasm32-wasi. The latter was used, because it compile using WASI API \footnote{https://wasi.dev/}, which is a system API that provides access to multiple operating system functionalities, such as access to the file system.

The resulting WASM file is then placed in a hidden folder in the home directory of a user, so it can later be loaded by the Go code. As the WASM code is used within a virtual machine (VM), it's independent from the operating system it will run on, so requires compiling only for one target.

\subsubsection{Calling WASM from Go using Wasmer}
\begin{figure}[h]
\centering
  \includegraphics[width=11cm, height=4.5cm]{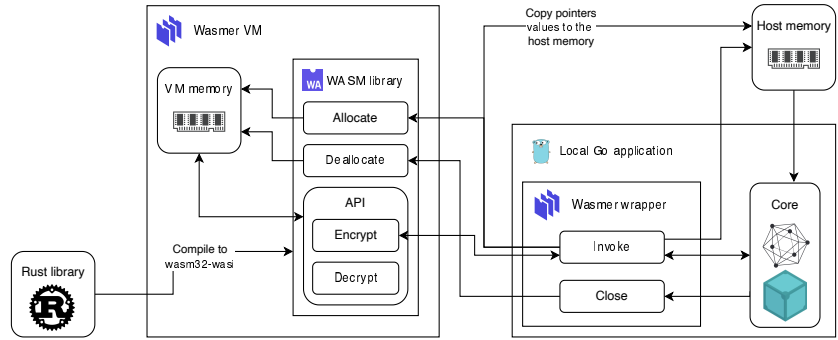}
  \caption{Connection of the WASM encryption library to the client application via the Wasmer library and the developed wrapper}
  \label{fig:encryptor_connection}
\end{figure}

In order to call WASM code a VM needs to be used. Wasmer library provides such VM, that can also be initialized from within Go code. The process of calling WASM code from the Go application is demonstrated in the Figure  \ref{fig:encryptor_connection}. This process consists of the following steps:
\begin{enumerate}
    \item Loading WASM code into a Wasmer VM
    \begin{enumerate}
        \item A directory on the host operating system, that will be accessible in the VM, need to be specified
        \item Optionally, standard output of the WASM library can be inherited.
    \end{enumerate}
    \item Invoking a function by specifying its name and the return type and passing arguments to it
\end{enumerate}

The communication between the Go code and WASM library and passing arguments for function invocation is happening using C types, which means that types like strings are not supported directly and need to be converted to corresponding C types. In case of a string being passed as an argument, it needs to be written to memory and end with a zero byte. The pointer to the first byte of this string is then passed to the invoked function as an argument.

As the host operating system memory is inaccessible for the VM, allocation and deallocation of memory need to happen within the VM itself. In order to facilitate the allocation and deallocation two dedicated Rust functions were developed as a part of id-based encryption library interface: allocate and deallocate. 

In case the called function requires a string as an argument, the allocation needs to be performed before passing the pointer to that string. The allocate function has to be called to allocate memory inside the VM. The memory is then accessed from the Go code and each byte of the string argument is written to the newly allocated memory. The pointer to the memory and the length need to be preserved in order to deallocate the memory, before the program terminates. The pointer to the first memory cell containing the string argument is then passed as an argument to the function that is being called.

\subsubsection{Wrapper code for Wasmer calls}
A wrapper code has been written to simplify invocation of WASM functions. The major simplification that this code provides is the ability to pass Go native-type argument to the wrapper, which then performs all the necessary processing and allocation, if needed. The pointers to string or array types as well as their lengths are stored, so when the program terminates, the memory is getting deallocated. 

Moreover, the developed wrapper code allows to pass simple numerical Go types (integers, floats, bytes, etc.) as pointers to the WASM library, so the changes that are happening to them when WASM functions run are also reflected in Go code, without the need to return anything. 

Additionally, the wrapper requires return type parameter argument (which is represented as an enumerator), when calling the invocation function through the wrapper. It uses the return type to case the return of WASM function to corresponding Go type. In cases when a pointer to a string is returned, the wrapper reads bytes from the VM memory until the zero byte and creates a Go string from it. The return type of the wrapper's invocation function is a generic \emph{interface\{\}}, which requires additional type casting. For example, in case the called function returned a pointer to a string, a Go string will be built from the pointer, but a user will still have to dynamically convert the returned value as it will be \emph{interface\{\}}.

\subsection{Smart contract}
\label{subsec:SmartContract}
The smart contract in Hyperledger Fabric allows to define assets that will be on the ledger. This paper defines an asset containing three fields: ID, Owner and CID. The code below shows the definition of an asset written in Go. 

\begin{lstlisting}[caption={The struct representing an asset on the Hyperledger Fabric ledger},captionpos=b,language=go]

type Asset struct {
	ID    string   `json:"ID"`
	Owner string   `json:"Owner"`
	CID   []string `json:"CID"`
}
\end{lstlisting}

The ID is a universally unique identifier (UUID) that is generated, when the new asset is created. The Owner is a string of hexadecimal numbers representing a public key of a user, who created the asset. The CID is an array if unique identifier that reference encrypted file chunks saved in IPFS. The references to the encrypted data chunks remain immutable, and can also be used for verifying if encrypted data chunks were manipulated (the hash of an encrypted data chunk is a unique value).

Additionally, the smart contract defines a list of functions for creating, deleting and updating assets. The implementation can be found on GitHub\footnote{https://github.com/ilyagrishkov/ib-self-encryption-smart-contract}.

\subsection{Self-Encryption work flow in a blockchain-IPFS based network}
The encryption and decryption process as well as interactions with the IPFS and the Hyperledger Fabric are orchestrated by a local application, which is a command line interface (CLI) tool written in Go. The prototype of the tool is accessible from GitHub\footnote{https://github.com/ilyagrishkov/ib-self-encryption}

Execution of any command starts with creating a new instance of a WASM wrapper and loading of the encryption library. When the command requires interaction with the Hyperledger Fabric, presence of the wallet, containing identity (which is necessary to enable interaction with the smart contract), is being checked. If the wallet is missing it's getting populated based on the certificates and keys of a user. When this preparation is done, the execution of the command starts.

At the end of the program execution the wrapper iterates over all allocated memory pointers and individually deallocates them.

\begin{figure}[h]
    \centering
  \includegraphics[width=12cm, height=9cm]{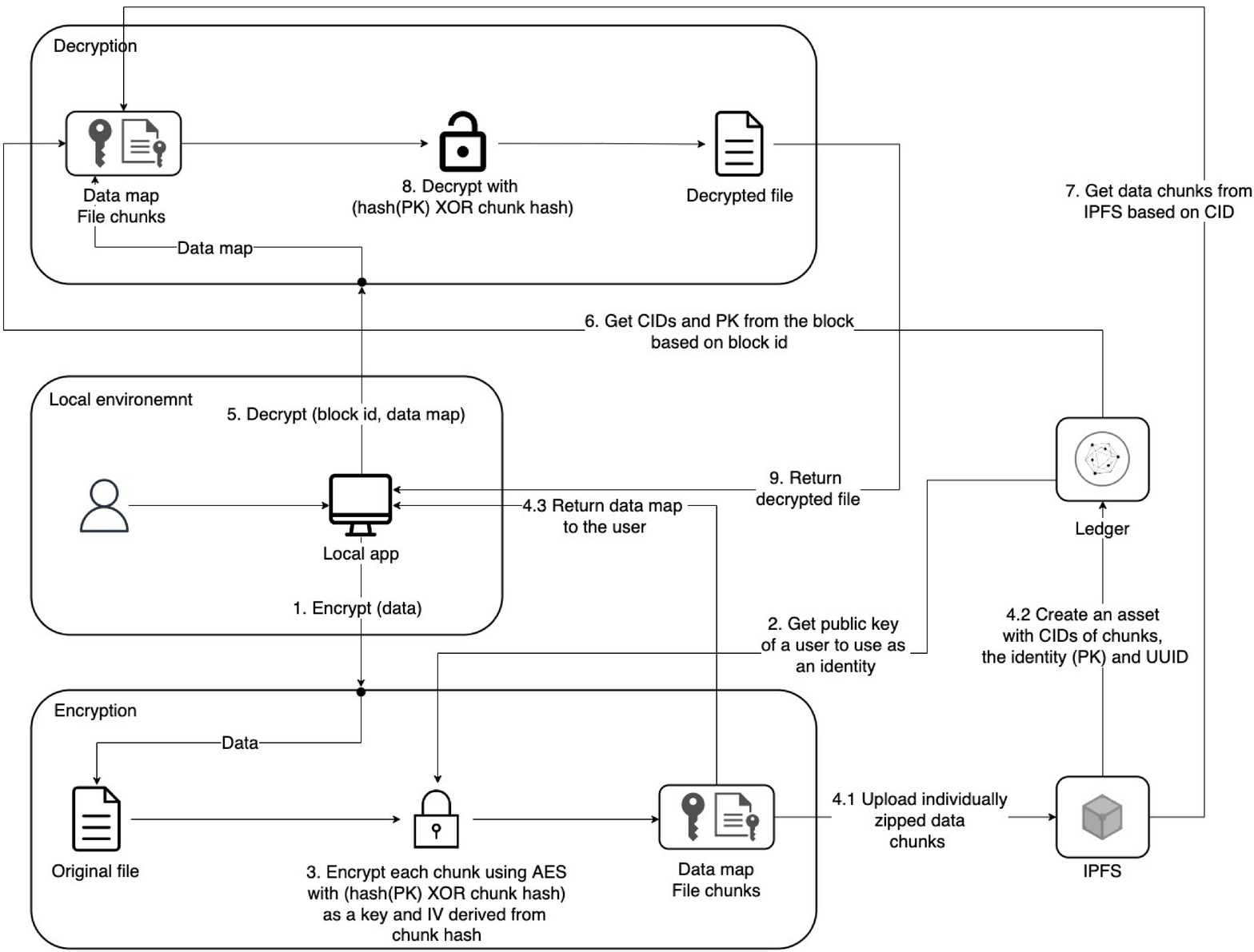}
  \caption{Work flow in the blockchain-IPFS based network when using self-encryption}
  \label{fig:workflow}
\end{figure}

There are two major parts of the system - encryption and decryption. Figure 3 demonstrates the workflow of both of them. 

\subsubsection{Encryption}
The first part, encryption, that deals with encrypting a file and uploading data to the Hyperledger Fabric starts when the following command is called:

\begin{lstlisting}[language=bash]
  $ ibse add [file] [key_output_path]
\end{lstlisting}

where \textit{ibse} is the name of the local app, \textit{file} is the absolute path to the file that needs to be encrypted, and \textit{key\_output\_path} is the absolute path to location where the key will be stored. 

The original file is is getting uploaded to the directory that was mapped during the VM initialization. From there it can be read by the WASM code. The encryption function is then called and the output is written to a new directory inside the mapped one. The output consists of multiple encrypted file chunks and a data map. The data map is moved to the location specified by the user and can later be shared via a secure channel. Each encrypted file chunk is being put into a zip archive to preserve their names, when uploading to the IPFS, and sent to the IPFS. The unique identifier, corresponding to each chunk (Content Identifiers or CIDs which are the hash values of the files) is returned. A smart contract function is then called that creates a new asset with all CIDs.

\subsubsection{Decryption}
The second part of the system, decryption, is invoked using the following command:

\begin{lstlisting}[language=bash]
  $ ibse get [block] [key] [destination]
\end{lstlisting}

where \textit{block} is the UUID of an asset in HLF blockchain that contains CIDs of encrypted chunks, \textit{key} is the absolute path to the data map, and \textit{destination} is the absolute path to location where decrypted file should be written. 

The UUID allows to identify an asset containing CIDs of encrypted file chunks. Each of the chunks is downloaded from the IPFS, unarchived, and written to the directory that is accessible from the VM. The data map is then copied to the same directory. After collecting all the necessary files for decryption, the decryption function is called and the restored file is written to a user-specified destination.  

\section{Results}
\label{sect:Results}
\subsection{Performance analysis}
Benchmarking of the system was done on the iMac 2019, 3,6 GHz 8-Core Intel Core i9 with 32 GB of memory running on MacOS 12.3.1. 

Benchmarking of the implemented id-based self-encryption scheme was done. As the encryption itself is not implemented in the same language as the rest of the project (the encryption is implemented in Rust and the rest of the project is in Go), the execution time can differ when Rust functions are called from Go compared to pure Rust execution time.

Files of sizes 100-, 250-, 500-, 750 kilobytes, 1 megabyte, 10-, 25-, 50-, 75-, and 100 megabytes were created for benchmarking both the pure Rust implementation as well as the WASM + Go implementations. Moreover, for this benchmark both the Rust code and the WASM library were optimized using maximum level of optimization provided by the Rust compiler.

\begin{figure}[h]
    \centering
  \includegraphics[width=9cm, height=7cm]{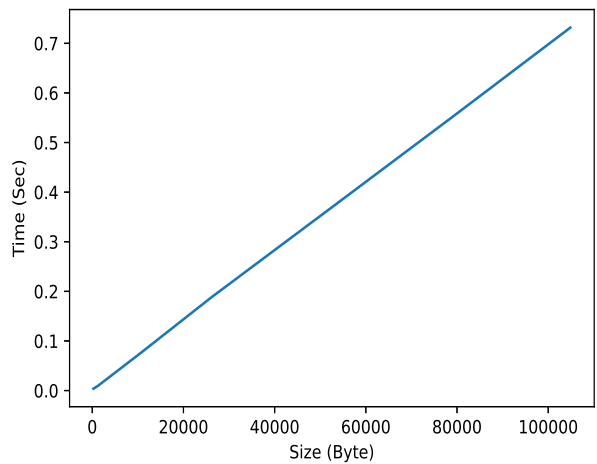}
  \caption{Dependence of the execution time of id-based self-encryption algorithm in pure Rust from the file size}
  \label{fig:workflow}
\end{figure}

The initial benchmark was performed on the encryption function only and was measuring execution time of the pure Rust implementation. Figure 4 shows the results of the benchmarking. 

The chart show near-linear dependence between the size of the file and the time it takes to encrypt it. This dependence can be explained by the fact that the most demanding computational is the AES 128 encryption process and with the increase of the file size, then number of chunks it is split to increases. Each chunk of the original file needs to be individually encrypted, hence the computation time grows linearly with the size of the file. 

As the encryption function execution time grows linearly due to the computational demand of the AES 128 and the hashing algorithms, the decryption process will be identical, because it uses the same algorithms for decryption.

In order to achieve more objective benchmark results, file of each size has been encrypted 100 times and the average calculated. In order to visualize execution time a chart in Python using MatPlotLib\footnote{https://matplotlib.org/} was created. The chart contains a 25-bin histogram, each representing density of a particular measurement. Following central limit theorem the distribution of the execution time measurements was assumed normal, so the mean and the standard deviation were calculated and the distribution plotted over the histogram. Figure 5 shows an example of combined charts for pure Rust and WASM + Go execution times, when encrypting 50 megabytes file. The blue histogram on the left-hand side shows results of the 100 measurements of the execution time of the Rust implementation; on the right-hand side - of the WASM + Go implementation.

\begin{figure}[h]
 \centering
  \includegraphics[width=9cm, height=7cm]{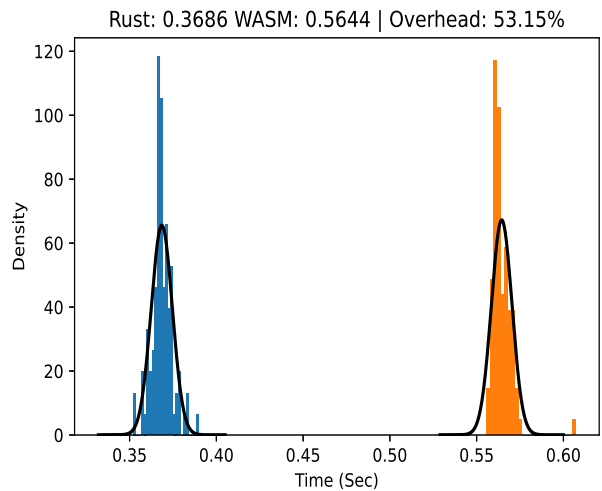}
  \caption{Run time distribution for 50MB file encryption using pure Rust and WebAssembly + Go implementations}
  \label{fig:workflow}
\end{figure}

The results of execution time measurements have for various file sizes are summarized in Table 1. The execution time shown in the table is the average number of seconds it takes a corresponding implementation to encrypt a file of a corresponding size. In addition to the average execution time, the overhead of the WASM + Go implementation is calculated for every pair of measurements.

\begin{table}[h]
\centering
\begin{tabular}{|l|rr|r|}
\hline
\multicolumn{1}{|c|}{\multirow{2}{*}{\begin{tabular}[c]{@{}c@{}}\textbf{File size} \\ \textbf{(Byte)}\end{tabular}}} &
  \multicolumn{2}{c|}{\textbf{Average execution time (Sec)}} &
  \multicolumn{1}{l|}{\multirow{2}{*}{\begin{tabular}[c]{@{}c@{}}\textbf{Overhead} \\ \textbf{(\%)}\end{tabular}}} \\ \cline{2-3}
\multicolumn{1}{|c|}{} &
  \multicolumn{1}{c|}{\textbf{Rust}} &
  \multicolumn{1}{c|}{\textbf{WASM + Go}} &
  \multicolumn{1}{l|}{} \\ \hline
100KB & \multicolumn{1}{r|}{0.0024} & 0.0042 & 75.89 \\ \hline
250KB & \multicolumn{1}{r|}{0.0038} & 0.007  & 84.51 \\ \hline
500KB & \multicolumn{1}{r|}{0.005}  & 0.0088 & 75.48 \\ \hline
750KB & \multicolumn{1}{r|}{0.0069} & 0.0117 & 71.4  \\ \hline
1MB   & \multicolumn{1}{r|}{0.0081} & 0.0139 & 71.29 \\ \hline
10MB  & \multicolumn{1}{r|}{0.0747} & 0.117  & 56.55 \\ \hline
25MB  & \multicolumn{1}{r|}{0.1885} & 0.2851 & 51.22 \\ \hline
50MB  & \multicolumn{1}{r|}{0.3686} & 0.5644 & 53.15 \\ \hline
75MB  & \multicolumn{1}{r|}{0.5492} & 0.8447 & 53.81 \\ \hline
100MB & \multicolumn{1}{r|}{0.7317} & 1.1201 & 53.08 \\ \hline
\end{tabular}
\caption{\label{tab:table-name}
Average execution time and overhead when encrypting files of different sizes using id-based self-encryption}
\end{table}

It is visible from the table that the overhead has a clear downwards trend (except the spike, when encrypting 250KB file). When the execution time of a WASM + Go encryption implementation is less than 0.01 seconds, the overhead falls in the range between 70\% and 85\%. When the execution time is longer than 0.1 seconds, the overhead goes down to 50\% - 55\% and stays in that range when the file size increases. Figure 6 demonstrates the overhead of WASM + Go encryption of file of different sizes.

\begin{figure}[h]
    \centering
  \includegraphics[width=10cm, height=7cm]{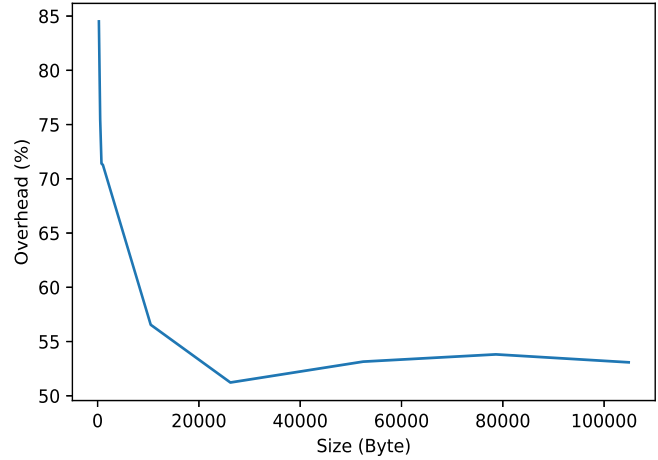}
  \caption{WebAssembly + Go implementation overhead measurements over for files of different sizes compared to pure Rust implementation}
  \label{fig:workflow}
\end{figure}

Such decrease in the overhead, when the execution time becomes longer is explained by the presence of the Wasmer library invocation overhead, which occurs every time a call is made to the Wasmer VM. When execution time itself is less than 0.01 seconds the invocation overhead is significant compared to the execution time. At the same time, when the execution time becomes longer, the overhead from invocation becomes insignificant, and measurements start to approximate real WASM VM overhead, which is around 50\% - 55\%. 

\subsection{Security analysis}
The designed app has multiple surfaces of attack. The IPFS nodes, where the encrypted file chunks are stored can attacked. Also, an adversary can be gain unauthorized access to the ledger with references to file on the IPFS. Both of those possibilities are analyzed below.

The security of IPFS nodes (assuming the encrypted file chunks were stored individually on multiple nodes) can be compromised, in which case encrypted files will be leaked to the malicious user. As encrypted file chunks have been stored on different nodes, the probability that all of them being compromised is negligible and should not be considered. Additionally, individual files do not have any link to each other, so matching multiple encrypted chunks, that are needed for successful decryption, is not computationally feasible. 
The data map containing the keys for the decryption, was stored locally by the user, who encrypted the file. Without the original keys, the decryption of the self-encrypted data is computationally not feasible \cite{irvine2010self}. Moreover, as the proposed self-encryption is also related to the data owner or user identity, the decryption cannot be done until the identity is not provided. Thus the data ownership is also provided by the id-based implantation.



\section{Discussion}
\label{sect:Discussion}
The results show high security guarantees of the id-based self-encryption scheme, when used for encrypting data, stored on the Hyperledger Fabric blockchain. This allows to use the implemented prototype as a secure medium for saving a retrieving information from the ledger.

In future works full Go implementation of self-encryption should be compared with the design proposed in this paper.

It was also beyond the scope of this study to create a standardized benchmarking for WASM and Rust libraries. It can be done by using multiple sample programs that test specific properties of the programming language (e.g. efficacy of memory allocation and deallocation) or very computationally intensive programs \cite{gouy}. The objective could be running a containerized version of both libraries against a set of such programs and analyzing the run time.

Moreover, the study can be expanded by analyzing and comparing CPU and memory load of WASM and Rust libraries. Such benchmark could be also done using sample programs mentioned in the previous paragraph. 

\section{Conclusion}
\label{sect:Conclusion}

In this study a new approach to storage of files on the Hyperledger Fabric blockchain was presented. The demonstrated approach allows for secure storage of data in a decentralized way, with ability to preserve the original file ownership and also information about the person, who encrypted it. This approach can be used where high security and trust in the integrity of data stored on the ledger is need. The prototype uses Rust implementation of id-based self-encryption that is compiled to WebAssembly and invoked from Go code. 

Additionally, the study demonstrates relatively low overhead (the overhead is around 55\%) and high performance level of WebAssembly library integration with the Go code base, compared to the pure Rust implementation. Relatively low overhead of WASM creates possibilities for developers to use WASM integration with Go and other languages that support Wasmer library as a cross-platform solution that allows to achieve high degrees of performance, while also being deterministic. 

The wrapper proposed in this work can also be used in Golang-based back-end applications that aim to use the cryptographic libraries deployed in Rust providing a more memory-safe execution.
\bibliography{samplepaper}

%
%
%
%




\end{document}